\documentclass{article}
\usepackage{amssymb}
\newcommand{\be}{\begin{equation}}
\newcommand{\ee}{\end{equation}}

\def\bea{\begin{eqnarray}}
\def\eea{\end{eqnarray}}
\def\bean{\begin{eqnarray*}}
\def\eean{\end{eqnarray*}}

\newcommand{\barr}{\begin{array}}
\newcommand{\earr}{\end{array}}

\newcommand{\bed}{\begin{displaymath}}
\newcommand{\eed}{\end{displaymath}}
\newcommand{\bal}{\begin{array}{ll}}
\newcommand{\eal}{\end{array}}

\def\bvec#1{\raise1.5ex\hbox{$\rightarrow$}\mkern-16.5mu #1}

\begin{document}

\centerline{\Large\bf  MEMOIRS OF AN EARLY STRING THEORIST\footnote{Invited Contribution to  ``The Birth of String Theory" Commemorative Volume}}
\vskip 1cm
\centerline{P. Ramond}
\vskip .3cm
\centerline{\em Institute for Fundamental Theory, Physics Department}
\vskip .1cm
\centerline{\em University of Florida, Gainesville, FL, 32611}
\vskip 2cm

\noindent I worked on String Theory over a period of five years during the First String Era, the most intellectually satisfying years of my scientific life. One of the early prospectors in the String Theory Mine, I was fortunate enough to contribute to the birth of this subject, which retains after these many years, its magical hold on our imaginations and expectations. 

\vskip 1cm
\noindent {\bf\Large Graduate School}
\vskip .2cm
\noindent{}
I was born and (mostly) raised in Neuilly/Seine, a suburb of Paris, where I attended Sainte Croix de Neuilly. After the  ``deuxi\` eme bachot"  in 1961,  I decided to spend a year with my family in New Jersey, where my civil engineer father had been working for a company that designed and manufactured concrete pipes. At his colleague's suggestions, I was enrolled at the Newark College of Engineering (NCE), known today as  the ``New Jersey Institute of Technology". There, I found spectacular teachers, especially Dr. Foster, and one year turned into four;  in 1965 I graduated from NCE with a Bachelor of Science in Electrical Engineering. The previous summer, I had worked at Bell Laboratories in Holmdel. My closest encounter with Physics had been a weird-looking antenna-like structure visible from the front of the Laboratory (neither I nor any of my fellow engineers knew its purpose).

Always interested in Physics\footnote{a cartoon from high school shows me with parabolic mirrors for which I had a special fascination}, I had studied the subject on my own while at NCE. Although I was awarded a Woodrow Wilson Fellowship in my senior year, my application to Princeton Graduate School in Physics was rejected (and wait-listed at Yale). Fortunately, I had also applied to Syracuse University where Peter Bergmann was teaching. With the recommendations of Professors Henry Zatzkis (a student of Bergmann), Mauro Zambuto,  and A. E. Foster, I was accepted, and soon afterwards, awarded a four-year National Defense Education Act (NDEA) Fellowship.

I had wanted to study General Relativity with Bergmann, but I was persuaded by Professor Ian McFarlane to switch to Particle Physics and join the group headed by E.C.G. Sudarshan.  Graduate School was not easy. My education as an engineer had left much to be desired, and the first year was a blurr of exams, coffee, and feelings of hopeleness.  Somehow I got through and E.C.G. Sudarshan took me on as his student. Before leaving on a sabbatical in India, he assigned me to prove that there is no  Spin-Statistics connection for infinite-component wave equations. That year, much work along the ``edge of the wedge" did not bring a definitive conclusion, but the task had become moot: upon his return from India, George told me that Ivan Todorov had found a counter-example!  After further studies of semi-leptonic Kaon decays in Sudarshan's new alternate theory of the Weak Interactions, I decided to switch advisors, and was accepted by A. P. Balachandran. There, I spent the rest of my graduate school safely inside the Mandelstam triangle, where crossing symmetry is manifest, trying to continue amplitudes from the $s$ to the $t$ and $u$ channels. 

I did not think I had done very well in Graduate School. My attraction to Physics comes from the beauty it suggests by providing simple answers to (apparently) complicated questions. Yet, in the heydays of the $S$-matrix approach, I found myself wandering in the complex plane, bleeding  from numerous encounters with its maze of poles and cuts. Not surprisingly, I did not get a job offer until late in the season (March 1969).  Bob Wilson, the director of the National Accelerator Laboratory (NAL), the High Energy Collider under construction in the western suburbs of Chicago, had decided to form a small theory group. I was to be one of five junior theorists,  ``The NAL Fives". It was a gamble, as the group had no Senior Theorists, but it was an opportunity to pursue my dream, and I became employee number $662$. NAL, known today as FermiLab, has been the site of many fundamental discoveries, most notably the top quark.  

\vskip 1cm
\noindent {\bf\Large Trieste}
\vskip .2cm
\noindent{}
Bal had kindly secured me a three-month appointment in the summer of 1969 at Professor Abdus Salam's ICTP in Trieste. It was a turning point in my scientific career. In Trieste I met Jean Nuyts with whom I had collaborated while a student. Jean and Hirotaka Sugawara were contemplating the beauties hidden behind the four-point Veneziano amplitude. It had recently been generalized to include many external legs, and intriguing  regularities were emerging from the factorization of the amplitudes. I was soon hooked, and the three of us decided to extract the three-point vertex from the amplitudes. We used tensor methods, and by the end of the summer, had succeeded in finding a (very ugly) expression. We were about to publish when the paper by S. Sciuto arrived, where the same vertex was derived using the harmonic oscillators techniques of Fubini, Gordon and Veneziano. His expression was so much simpler than ours that we did not publish our results. In Graduate School, I had already encountered the simplicity of Dirac's creation and annihilation operators, not only in quantization but also in Schwinger's treatment of $SU(2)$. They were clearly the window into the structures behind the Dual Models. 

It is also in Trieste that I had the first glimpse of my hero, P. A. M. Dirac. A tall and lonely man, he stood aside from all, and gave the impression of being in such deep thoughts, that no-one, well almost no-one,  dared to intrude. One  lecture found Dirac fast asleep in the second row, and me a few rows back,  musing about the smallness of such a great man's head! In the middle of the lecture, a woman strode down the seminar room, tapped Dirac on the shoulder and said, to my horror: ``Come on Paul, you can sleep at home!". He followed her meekly; this was my introduction to Margit Dirac.  

\vskip 1cm
\noindent {\bf\Large NAL Theorist}
\vskip .2cm
\noindent{}
I came back to America and NAL, determined to learn more about these new techniques. In the middle of the Atlantic on the liner ``France", I was studying a Fubini-Veneziano preprint in the ship's reading room. Imagine my surprise when I saw  the {\em same} preprint on another desk! Its owner was nowhere to be seen, and it was annotated in a different hand. I eagerly waited for the reader. This is how I met  Andr\' e Neveu who was on his way to Princeton University. We decided to keep in touch about each other's progress. My wife Lillian and I were amazed when we asked Andr\' e to join us at the bar, and found him already in his pajamas at 8 p.m.! I wondered if all physicists were like that.

NAL was more like a summer camp than a laboratory. It was a collection of houses away from the construction site. The five theorists were housed in one house, experimentalists in another\footnote{See http://bama.ua.edu/$\sim$lclavell/Weston/ for Lou Clavelli's   wonderful account of these early NAL days }.  The director's complex was a bunch of houses put together. Over lunch, Bob Wilson informed us that eventually theorists should be housed in open areas without doors so as to increase communications. I objected, and remember the silence that followed! 

One of the five theorists was Lou Clavelli whom I had already met when he had spent a month in Trieste.  David Gordon had worked with Fubini and Veneziano in developing the oscillator formalism, and I looked forward to collaborating with him.  David Gordon and I wrote the {\em first} theory paper out of NAL (THY-1), a not very memorable attempt to include fermions into the Veneziano model.

Jim Swank, a student of Nishijima, and Don Weingarten, a student of Bob Serber completed the quintet. David, Lou, and I shared an interest in studying the Dual Resonance Model amplitudes. I did not yet know about the string connection. In fact, when I told Lou that the mass spectrum looked familiar and reminded me of something I had come across, his response was: ``Nambu says it is a string". Lou had been Professor Y. Nambu's graduate student at Chicago, and he soon arranged a meeting over lunch with Nambu at the Quadrangle Club. There the great man gently encouraged us in our  studies, asking probing questions with no easy answers that left me totally impressed. He even treated us to the lunch! Brilliance and humility, a combination of traits not often seen in physicists. 

 We all visited the University of Wisconsin where we met, among others, Bunji Sakita and Miguel Virasoro. They both seemed off-scale, and so way ahead of us in their understanding of Dual Models. In particular, Miguel had just constructed a set of operators which had the potential to decouple the negative norm states found in the amplitudes\footnote{Decoupling was later proved by Brower, Goddard and Thorn}, but only when the lowest particle was a tachyon and the first excited state massless.  He had spent a lot of time trying to get around these ``unphysical" constraints, and hesitated to publish although all who saw his work agreed it was too beautiful not to. 

Nambu was the only senior theorist with whom we had scientific discussions. Many senior theorists came to NAL to gauge the progress and visit the site. The ``NAL Fives"  were given guide duties: escorting visiting theorists up a Silo near the ring (then under construction), taking them to lunch, and setting up a Colloquium/Seminar. For our troubles, we got a free lunch at a restaurant off-site. Very few showed any interest in what we were doing.

Andr\'e and I had agreed to keep in touch and report on our mutual progress. I soon received a preprint of Andr\' e Neveu and Jo\" el Scherk who had successfully isolated the infrared divergence of the planar one-loop dual amplitude. It was a feat of mathematical physics, that only those with French training could have achieved: they used the Jacobi imaginary transformation. I invited them to NAL. This is how I met Jo\" el and his first wife. I was impressed by the fact that they had met at Club Med, not the usual hangout of physicists!  Jo\" el was as quiet and introspective as Andr\' e was exhuberant and flashy. They came across as intellectually brilliant, and I took to both right away. 

Lou and I found it difficult to work with David Gordon, so we continued our studies of the Dual Resonance Models on our own. It took us a long time before we published our first paper (May 1970). In it we put on firm group-theoretical basis the vertex  introduced independently by Nambu and Fubini and Veneziano, using the $SU(1,1)$ operators found by Gliozzi, Chiu-Matsuda-Rebbi, and Thorn. From there we understood that fixing the (not yet known as) conformal weight of the vertex was the same as a (free) equation of motion for the emitted particle. We quickly set about to find the vertex for the first excited state. It had never been written down before and this was to be our first truly original work.  

\vskip 1cm
\noindent {\bf\Large Aspen}
\vskip .2cm
\noindent{}
Early in 1970, Bob Wilson, bless his soul, decreed  ``All theorists must go to Aspen"! As Bob was the boss, we went, although none of us knew much about the Aspen Center for Physics. This was the best advice an experimentalist ever gave me! The Aspen Center for Physics was at the time a collection of two buildings, set near the music tent at the West end of Aspen, Colorado. The oldest, Stranahan Hall, was made of bricks; the second, Hilbert Hall,  was a wooden structure which had been donated to the Center by Bob Wilson after it was used for the 1968 NAL Summer Study. 

It was a wonderful stay. The first day in Aspen, Don Weingarten and I went to dinner at Guido's, where Don confiscated a sign which proclaimed something like ``no hippie barefoot people allowed in the restaurant"! Little did we know that the same Guido was known for having chased after hippies with his shotgun! Don (I was told) had achieved early notoriety during student riots by having  his picture taken, sitting in  the President of Columbia University's armchair, and smoking the President's  cigars! The town was in the afterglow of the hippie era, and my days were spent playing volleyball in Wagner Park, listening to music outside the music tent in the late afternoons, and in other non-scientific activities. In my spare time, I started thinking about the particle spectrum that had been extracted from the Dual Amplitudes. People had already found some sort of position operator $Q_\mu(\tau)$ which appeared in the vertex, and its derivative was like a generalized momentum $P_\mu(\tau)$. Indeed, if one pursued the analogy further, the inverse propagator looked like the square of that generalized momentum. This led me to think of a {\em Correspondence Principle} by which simple notions of point particles were related to Dual Models. At last, a glimpse of simplicity! Was it the altitude, was it the easy-going atmosphere at the Center that enabled me to view the problem with a different eye? I will never know but that summer stay changed my perspective forever.  

Back at NAL, Lou and I spent the rest of the summer refining our results on the excited vertex, and abstracting from it general rules for the construction of Dual Amplitudes, which we would call today conformal theory. We submitted that work to the {\em Physical Review} in late September. Although interested, Lou did not share my enthusiasm for the Correspondence Principle and in October, I sent the paper on the Correspondence Principle (applied to bosons only) to {\em Physics Letters}. 

In the process, I realized that it could be applied to the Dirac equation: all I had to do was to generalize the Dirac matrices.  To my surprise, this led to an algebra of a kind I had never seen: it contained both  commutators and anticommutators, and was in essence  the square-root of the Virasoro algebra. Tremendously excited, I barely ate and drank for weeks, as every derivation brought more conceptual clarity and more questions. There were some odd things; the generalization of the Dirac matrices led to fermionic harmonic oscillators $b^{(n)}_\mu$ and $b^{(n)\dagger }_\mu$ with space-time four-vector indices, but they came with their own operators, $F_n$ of the right structure to decouple the negative norm states. I also realized that it was a truly novel algebraic structure, since I could now take the square root of {\em any} Lie algebra\footnote{At Yale, I approached the great algebraist, Professor N. Jacobson in the Mathematics Department. I showed  him my algebra of commutators and anticommutators, asking if he had seen anything like this. I must not have made a convincing case since he replied in the negative and kindly ushered me out of his office}. I explained my results to Don Weingarten over lunch and he simply said: ``You are set for life". Of course I did not believe him.

Lillian, an electrical engineer, and NCE alumna, had left for a six week stay at a Western Electric's Training Center near Princeton, and I joined her for a few days, sometimes in late October, as I recall. I took advantage of the trip to visit Y. Nambu who was spending the Fall at the Institute for Advanced Study. There I also met Jamal Manassah, Satoshi Matsuda, and Mike Green, and sketched my ideas on including the fermions via the Correspondence Principle. I was visiting Lillian (as the only male spouse!) and had no time to see Andr\' e at the University.

Back at NAL, a host of distractions caused me to start losing focus. First, Ned Goldwasser informs us in early November that our appointments at NAL will terminate in 1971 (except for David Gordon's). This came as a big surprise: when hired, we had been  told to expect longer appointments because of the special circumstances, such as the lack of a senior theorist.  This was devastating news for us all. My own work had not been manifestly relevant  to NAL, but Lou and the others had always kept close to experiments, and worked hard in NAL-sponsored workshops.  Bob Wilson was listening to senior US theorists who then had no sympathy for Dual Models/String Theory. We all became anxious about finding another job. My last experience had been nerve-wracking, and I was soon freaking-out. Secondly, the Correspondence Principle paper was rejected by {\em Physics Letters}. I argued but to no avail, and in December I withdrew the paper (Maurice Jacob, the Editor, told me years later that the rejection letter was not so strong, and that it could have been accepted with corrections, but I did not know the rules of the game). 

Job applications took up most of my time, preventing me from writing up the fermion paper, and it was not until the week after Christmas that I sent it to the {\em Physical Review}, as well as to Andr\' e and others. At the same time, I resubmitted  the Correspondence Principle paper, this time to {\em Il Nuovo Cimento}, and went back to the business of job-hunting. 

In those days, postdoctoral positions were on the whole awarded at the January Meeting of the American Physical Society in New York. This January slave market was a depressing affair for Lou and I: our theoretical teeth and muscles were examined, but with no apparent outcome. The establishment had no time for Dual Resonance Models/String theory, and showed minimal interest in what we were doing. 

Then I got lucky, through a set of fortuitous circumstances orchestrated by Lou Clavelli. Before joining NAL, Lou had post-doctored at Yale, where he had met his future wife, Estelle. As he was to visit his in-laws after the New York meeting, he suggested I join him and use the opportunity to visit the Yale High Energy Theory group, to which I had already applied. That Friday the Yale physicists, Sam McDowell, Charlie Sommerfield, John Harte, and Dick Slansky (Feza G\" ursey was out of town) asked me to give an impromptu seminar, my first seminar on the Dual Dirac equation. The same night we flew back to Chicago, and three days later, Charlie Sommerfield offered me a one-year instructor position at Yale with the possibility of a second year! Finally, an early job offer! After talking with Lillian to discuss our next move, I called Charlie just a few hours later, accepting his offer! He told me later he was surprised by the speed of my acceptance, but then again I did not know the rules.

Three months after my fermion paper, I received a preprint by Andr\' e Neveu and John Schwarz who proposed the Dual Pion Model, a generalization of the Veneziano model. It had not external fermions, but included fermionic ladder operators with space-time vector indices, just like my model of fermions,  with the same name but different labeling. The underlying algebraic structure they proposed was the same as  mine.  Andr\' e  told me later when I visited him in Princeton (when I first met John Schwarz), that they had indeed been motivated by my paper: they had introduced a Yukawa interaction into my model, computed the amplitude with two fermions and an arbitrary number of bosons. They had great difficulties with the zero modes in the fermion sector (who didn't?), but they factorized the amplitude in the cross channel to extract the Dual Pion Model amplitudes. 

Their paper was  wonderful, but imagine my surprise when I found the fermion paper mentioned only towards the end of theirs. I remain to this day baffled by their lack of acknowledgment of the seminal role of my work.

In 1969, Nambu had suggested that the Veneziano amplitudes should be derivable from a string. Later, Nambu and independently Goto, had proposed an action for a relativistic string. It was not linear, and seemed impossible to quantize.  When Virasoro found his  decoupling coefficients, Nambu had interpreted them as evidence for his string picture, as the generators of the conformal algebra generated by the Fourier coefficients of the energy-momentum tensor of a two-dimensional theory spanned by the world sheet of a relativistic string. His point of view, while intriguing and interesting, was not widely appreciated, as it did not offer any computational advantages over the well-developed and fruitful amplitude approach.  

The Spring of 1971 was spent in adding electromagnetic interactions to the Dual Dirac equation, and in trying to find a meaning for the anticommutators. I understood the anticommuting operators as generators of transformations between bosons and fermions.  Such close kinship between fermions and bosons did not surprise me; it was natural in the original formulation of duality applied to the pion-nucleon amplitude, with its implied relations between its fermionic $s$-channel and bosonic $t$-channel. Unfortunately,  I became confused by the parameters of these transformations (I did not know about Grassmann numbers), and did not get very far.  

This symmetry between fermions and bosons was of course the first manifestation of a new type of symmetry, called (later) {\em Supersymmetry}.  It was first found in String Theory, which has proven to be an incubator for many new ideas. 

Unfortunately, by that time I was physically and emotionally spent. I fell sick with high fever for several days (most unusual for me), and  was diagnosed with kidney stones, which were surgically removed. 

That experience was not all negative. Lillian and I befriended the NAL staff physician, Dr Cornell, quite a character who had jumped on D-Day with the Airborne in Normandy. Today his favorite chicken recipe,  ``Chicken \` a la Cornell" occasionally graces our table.  In the DuPage County Hospital,  I also became a friend of Jim Ward, a police sergeant from Wheaton, the town we lived in. I organized a visit to NAL (I had much practice by that time) for his younger son Scott, who later credited me for awakening his interest in Physics; he is today a physicist at NREL in Golden, Colorado, working on Solar Cells.  

I gave the second (and last) seminar on the dual fermions  at the University of Chicago where I found a sympathetic reception. However the outside world did not seem to show a similar interest.  My fortunes changed when Professor Stanley Mandelstam (in whose triangle I had toiled as a graduate student) visited NAL. I excitedly told him about the fermions, expecting a reaction; there was none. Later that day, I pressed him for his opinion, and still remember his answer for its honesty:  ``You claim to have done something that many have tried to do, including my colleagues at Berkeley. I need time to assess it".  True to his word, Stanley, together with Nambu, proved to be a most generous advocate of my work. 

Also, Lou, Andr\' e  who had stopped by NAL on his way to Berkeley, and I discussed the $F_n$  as decoupling conditions (never published). 

In the Summer, I was invited to lecture at the Boulder Summer School on the construction of dual amplitudes for vertices of arbitrary spin that Lou and I had developed. Many visiting senior theorists came through NAL, although the local theorists were largely ignored. There were of course exciting developments taking place elsewhere: Ben Lee, fresh from the Amsterdam Conference, brought the news of a young Dutch theorist who had proved that some massive Yang-Mills theories  are renormalizable. This rang a bell because at Syracuse, Joe Schecter had suggested I read Weinberg's 1967 paper, ``A Theory of Leptons",  although it was believed at the time that one could not calculate anything beyond tree level. I wondered if the tachyon of the Veneziano model was a signal of spontaneous breaking, but then of what? In the Fall of the same year, at a conference in Rutgers, I tried to interest my fellow theorists in this problem, but to no avail. This was where I learned from Claude Lovelace that cuts disappear from the twisted one-loop diagram when the number of space-time dimensions is twenty six. None of us understood the significance of this result.

I joined Yale in Fall 1971 as an Instructor. I was quite taken by the friendly atmosphere of the High Energy Theory group headed by the Turkish physicist Feza G\" ursey. He and his wife Suha provided a great intellectual and humane cocoon into which Lillian and I were readily accepted.  Yale was a wonderful place, and my colleagues there have remained life-long friends.   

As Yale Instructor and later Assistant Professor, I continued to work in string theory. My student Michael Kalb and I understood the fundamental role of antisymmetric tensors in theories of extended objects, using ideas from Action-at-a-Distance theories where these fields are linked to mutidimensional world-sheets. Today the two-form $B$-field is a mainstay of string theories. Little did I know that their study would presage branes and so many other wonderful developments!  By the time of the 1974 London Conference, I had started working on Exceptional Groups and their applications to Grand-Unified Theories. My last string publication of the First String Era was with another Yale student, Charles Marshall, on the covariant formulation of String Field Theory. 

The decision to leave string theory was primarily driven by the community's lack of interest and dearth of jobs in this wonderful subject; at the same time Lillian and I had to provide for our two children. Also, I was also lured away by the intellectual promises of Feza's application of unusual algebraic structures to Physics. I was, needless to say, very pleased when many years later, Exceptional Groups appeared in the Heterotic String. 

\vskip .5cm
\noindent {\bf\Large Acknowledgments}
\vskip .2cm
\noindent I am grateful to the organizers of the ``The Birth of String Theory",  Filippo Colomo and  Paolo Di Vecchia, for giving me the opportunity to present my recollections. I also wish to thank Sudarshan Ananth, Lars Brink and Lou Clavelli for reading of the manuscript and their many helpful suggestions.  This work, supported by the Department Of Energy Grant No. DE-FG02-97ER41029, was begun at the Aspen Center for Physics which I thank for its hospitality. 

My earlier scientific presentation at the Galileo Galilei Institute for Theoretical Physics can be found on its website, http://theory.fi.infn.it/colomo/string-birth/. 

\end{document}